\documentclass{article}

\usepackage[latin1]{inputenc}
\usepackage[dvips]{graphicx}
\usepackage{amsmath}
\usepackage{amsfonts}
\usepackage{amssymb}
\usepackage{amsthm}
\usepackage{enumerate}
\usepackage{mathrsfs}

\newtheorem{theorem}{Theorem}
\newtheorem{lemma}[theorem]{Lemma}

\newtheorem*{corollary}{Corollary}

\begin{document}

\title{Composability in a certain family of entropies}
\author{Henrik Densing Petersen\footnote{Personal contact information: Henrik Densing Petersen, Melissehaven 23, 2.th., 2730 Herlev, Denmark. Phone: +45 61717642. Fax (at the institute): +45 35320704} \\ Department of Mathematical Sciences \\ University of Copenhagen \\ m03hdp@math.ku.dk}
\maketitle

\begin{abstract}
It is shown that the Tsallis entropies are the only entropies of the form $H(P)=-\sum_i f(p_i)$, with suitable assumptions on $f$, satisfying the condition of composability.
\end{abstract}


\section{Introduction}
We define the $f$-entropy as in \cite{topsoe06} (see also \cite{topsoe04} for some basic properties), namely, we demand that $f$ is a real-valued analytic and strictly convex function on $[0,\infty )$, that it satisfies $f(0) = f(1) = 0$, and lastly, that it is normalized so that $f'(1)=1$. The $f$-entropy is then defined as
\begin{equation}
H_f(P) = -\sum_{i\in A} f(p_i),
\end{equation}
where $P=(p_1, p_2, \dots)$ is a probability distribution on $A$. Unless explicitly stated otherwise we shall assume that $A=\mathbb{N}$.

Simple considerations show that for any probability distribution $P$, $0\leq H_f(P)<-f'(0)$ and that $\sup\{H_f(P)\vert P\}=-f'(0)$. Thus $g$ in (\ref{eq:intro}) need only be defined on $[0,-f'(0)[\times [0,-f'(0)[$.

The purpose of this paper is to show that the Tsallis entropies are the only $f$-entropies that are composable, i.e. satisfy the equation
\begin{equation}
\label{eq:intro}
H_f (P\otimes Q) = g(H_f(P), H_f(Q))
\end{equation}
for some function $g$ and all probability distributions $P,Q$. The possible forms of the function $g$ for general measures of entropy is considered in the articles \cite{abe01, johal04}. The relation between non-additive and additive entropies is explored in \cite{johal03}.

The Tsallis entropies were first considered in the mathematical literature in \cite{havrda} and then later (and independently) in statistical physics by C. Tsallis in \cite{tsallis88}. They are defined for any $q>0$ as the $f$-entropy associated with the function
\begin{equation}
f(x)=\frac{x-x^q}{1-q}. \nonumber
\end{equation}
It is not hard to see that the Tsallis entropies in fact satisfy the equation
\begin{equation}
\label{eq:tsallis}
H_q(P\otimes Q) = H_q(P) + H_q(Q) + (1-q)H_q(P)H_q(Q).
\end{equation}
Our first theorem is the converse: Any $f$-entropy satisfying an equation of this type must be a Tsallis entropy. Having proved that, we can extend the result to cover $f$-entropies satisfying the more general equation (\ref{eq:intro}).

The paper \cite{Hotta} addresses the same issue as here. Let us briefly describe how the present work differs in outlook. In \cite{Hotta}, the "non-obvious" assumptions needed are the existence of $f'(0)$ as well as either that of $f$ being a sum $f(x)=\sum_{\textrm{some } \alpha'\textrm{s} \ge 1} x^{\alpha}\phi_{\alpha}(x)$ or an assumption of analyticity on $g$.

Assuming the existence of $f'(0)$ seems rather limiting, recalling that for the shannon entropy, $(x\log x)'$ indeed does not exist for $x=0$. As for the other two assumptions, it is not at all a priori clear how restrictive they are in physical terms.

It seems much more natural to take an important physical property of the entropy, in our case concavity (which is equivalent to the convexity of $f$), and starting from there. This gives a precise result, which seems to have exactly the right relevance for physical applications.

\section{Composability of $f$-entropy}
\begin{theorem}
\label{thm:converse}
If the $f$-entropy satisfies
\begin{equation}
H_f(P\otimes Q) = \alpha H_f(P) + \beta H_f(Q) - \gamma H_f(P)H_f(Q),
\end{equation}
for constants $\alpha, \beta, \gamma$ and all probability distributions $P,Q$, then $\alpha = \beta = 1$, and
\begin{equation}
\label{eq:f1}
f(xy) = xf(y)+yf(x)+\gamma f(x)f(y)
\end{equation}
for all $x,y \ge 0$.
\end{theorem}
For a more general version of this theorem, see for example \cite{abe00}. We give an elementary
\begin{proof}
Plugging in the definition of the $f$-entropy, we get that for all probability distributions $P=(p_i)$, $Q=(q_j)$,
\begin{equation}
\label{eq:fform}
\sum_{i,j} f(p_iq_j) = \alpha \sum_if(p_i) + \beta \sum_jf(q_j) + \gamma \sum_{i,j} f(p_i)f(q_j).
\end{equation}
Letting $Q$ be deterministic yields $\alpha=1$ and $P$ deterministic yields $\beta=1$. For uniform distributions $P=U_m$ and $Q=U_n$, we then find from (\ref{eq:fform}) that
\begin{equation}
\label{eq:otte}
f(\frac{1}{mn}) = \frac{1}{m}f(\frac{1}{n}) + \frac{1}{n}f(\frac{1}{m}) + \gamma f(\frac{1}{m})f(\frac{1}{n}),
\end{equation}
that is, $f$ satisfies (\ref{eq:f1}) for all $x=\frac{1}{m}$, $y=\frac{1}{n}$. We seek to extend this to all rational numbers in $[0,1]$, which then extends to all $x,y \in [0, \infty )$ by the analyticity of $f$.

Let $q\in [0,1]\cap \mathbb{Q}$ and choose $a,m\in \mathbb{N}$ such that $q+a\frac{1}{m}=1$. Applying (\ref{eq:fform}) to $P=(q,\frac{1}{m}, \dots ,\frac{1}{m}, 0, \dots )$ and $Q=U_n$, we get
\begin{equation}
nf(\frac{q}{n}) + anf(\frac{1}{mn}) = f(q) + af(\frac{1}{m}) + nf(\frac{1}{n}) + \gamma \left(f(q)+af(\frac{1}{m})\right)nf(\frac{1}{n}), \nonumber
\end{equation}
and so using the above result on the term $anf(\frac{1}{mn})$ and remembering that $\frac{a}{m}=1-q$, we get
\begin{equation}
\label{eq:ti}
f(\frac{q}{n}) = qf(\frac{1}{n}) + \frac{1}{n}f(q) + \gamma f(q)f(\frac{1}{n}).
\end{equation}
Finally, letting also $r\in [0,1]\cap \mathbb{Q}$ and choosing $b,n\in \mathbb{N}$ such that $r+b\frac{1}{n}$, we apply (\ref{eq:fform}) to $Q=(r, \frac{1}{n}, \dots , \frac{1}{n}, 0, \dots )$ and $P$ as above, getting
\begin{eqnarray}
&&f(qr)+bf(\frac{q}{n})+af(\frac{r}{m})+abf(\frac{1}{mn}) = f(q)+af(\frac{1}{m})+f(r)+bf(\frac{1}{n})+ \nonumber \\
&&+ \gamma\left( f(q)+af(\frac{1}{m})\right) \left( f(r)+bf(\frac{1}{n})\right). \nonumber
\end{eqnarray}
Using (\ref{eq:ti}) and (\ref{eq:otte}) on the left-hand terms we get that
\begin{equation}
f(qr) = qf(r)+rf(q)+\gamma f(q)f(r).
\end{equation}
Since this holds for all $q,r \in [0,1]\cap \mathbb{Q}$, the theorem follows by analyticity of $f$.
\end{proof}
That this does indeed provide a converse to (\ref{eq:tsallis}), is shown by the following
\begin{corollary}
If the $f$-entropy $H_f$ satisfies the conditions of the theorem, then $H_f$ is a Tsallis entropy: $H_f = H_q$ for some $q>0$.
\end{corollary}
\begin{proof}
In \cite{topsoe06} it is shown that if $f$ satisfies the factorization property
\begin{equation}
\sum_i \left(q_if(\frac{p_i}{q_i}) - f(p_i)\right) = \sum_i \xi(p_i)\zeta(q_i), \nonumber
\end{equation}
where $\xi, \zeta$ are analytic functions on $[0, \infty [$ satisfying $\xi(1) = 1$ and $\zeta(1) = 0$, then the $f$-entropy is a Tsallis entropy.

Well, we have by the theorem that
\begin{equation}
q_if(\frac{p_i}{q_i}) - f(p_i) = (p_i+\gamma f(p_i))q_if(\frac{1}{q_i}),
\end{equation}
hence the result follows.
\end{proof}

Before we can tackle the general problem, we need to note an important property of $f$-entropy. Namely, let $Q=(q_i)_{i\le N}\in M_+^1(N)$\footnote{For $N\in \mathbb{N}$ we write $M_+^1(N)$ for the set of probability distributions on $\{ 1, \dots ,N\}$.} with say, $0<q_1<q_2<1$, and consider also the distribution $Q_{\varepsilon} = (q_1-\varepsilon, q_2+\varepsilon, q_3, \dots , q_N)$. If $\varepsilon$ is sufficiently small and negative - more precisely, if $\frac{1}{2}(q_1-q_2)\le \varepsilon <0$, we have that the entropy of $Q_{\varepsilon}$ is strictly greater than the entropy of $Q$. Similarly, if $\varepsilon$ is sufficiently small and positive - more precisely, if $0<\varepsilon\le q_1$, the entropy of $Q_{\varepsilon}$ will be strictly smaller than that of $Q$. This follows by the strict convexity of $f$.

Together with the continuity of $H_f$ on $M_+^1(N)$, this implies that the mapping $\varepsilon \curvearrowright H_f(Q_{\varepsilon})$ is a homeomorphism of the interval $[\frac{1}{2}(q_1-q_2), q_1]$ onto the interval $[H_f(0, q_2+q_1, q_3, \dots, q_N) , H_f(\frac{1}{2}(q_1+q_2), \frac{1}{2}(q_1+q_2), q_3, \dots, q_N)]$.

\begin{lemma}
\label{lma:to}
Let $0<a<b<-f'(0)$. Then there exists $N\in \mathbb{N}$, and $\tilde{q}_1, \tilde{q}_2$ with $0<\tilde{q}_1<\tilde{q}_2<1$ such that
\begin{equation}
[a,b]\subseteq \{ H_f(Q)| Q=(\tilde{q}_1,\tilde{q}_2,q_3,\dots ,q_N)\in M(\{1, \dots, N\}) \}.
\end{equation} 
\end{lemma}
The point of this is, that we can fix $N, \tilde{q}_1, \tilde{q}_2$ once and for all, and still have enough probability distributions to \lq\lq calculate with $g(x,y)$''  with $x,y\in ]a,b[$.
\begin{proof}
Since the sequence $(-nf(\frac{1}{n}))$ is increasing and converges to $-f'(0)$ (indeed, by definition $f'(0)=\lim_{n\rightarrow \infty} nf(\frac{1}{n})$), all $N$ sufficiently large will satisfy
\begin{equation}
b<-(N-1)f(\frac{1}{N-1}) = H_f(U_{N-1}).
\end{equation}
So, from the remarks above, it is clear that by choosing $N$ sufficiently large and $\tilde{q}_1+\tilde{q}_2 = \frac{1}{N-1}$, with $0<\tilde{q}_1<\tilde{q}_2$, we get
\begin{equation}
H_f\left( \tilde{q}_1, \tilde{q}_2, 1-\tilde{q}_1-\tilde{q}_2\right) < a < b < H_f(U_{N-1}) < H_f\left( \tilde{q}_1, \tilde{q}_2, \frac{1}{N-1}, \dots, \frac{1}{N-1}\right).
\end{equation}
The lemma then follows since we can get the right-hand entropy from the left-hand one by continuously pulling probabilities together, as in the remarks above.
\end{proof}

\begin{theorem}
\label{thm:composability}
Assume that the $f$-entropy $H_f$ satisfies the condition of composability as in eq. (\ref{eq:intro}) for some function $g$ defined on $[0,-f'(0)[\times [0,-f'(0)[$. Then $g$ has partial derivatives on $]0,-f'(0)[\times ]0,-f'(0)[$, and indeed
\begin{equation}
g(x,y) = x + y + \gamma xy, \quad x,y \in [0, -f'(0)[,
\end{equation}
for some constant $\gamma$.
\end{theorem}
\begin{proof}
We want to show that on all sets $]a,b[\times ]a,b[$ with $0<a<b<-f'(0)$, the partial derivatives exist and satisfy
\begin{equation}
\frac{\partial}{\partial x}g(x,y) = A(y), \quad \frac{\partial}{\partial y} g(x,y) = B(x). \nonumber
\end{equation}
The theorem then easily follows.

So, for given $a,b$ choose $N, \tilde{q}_1, \tilde{q}_2$ as in the lemma. Let $x,y\in ]a,b[$ and choose $P,Q=(\tilde{q}_1, \tilde{q}_2, q_3, \dots, q_N) \in M_+^1(N)$ such that
\begin{equation}
H_f(P)=x, \quad H_f(Q)=y. \nonumber
\end{equation}
If $y_n\rightarrow y$ for some sequence $(y_n)$, we can assume\footnote{By leaving out finitely many of the $y_n$'s if we have to, and using that the mapping $\varepsilon \curvearrowright H_f(Q_{\varepsilon})$ is a homeomorphism of intervals as remarked above.} that there exists a sequence $(\varepsilon_n)$ such that for all $n$
\begin{equation}
y_n = H_f(Q_{\varepsilon_n}) = (\tilde{q}_1-\varepsilon_n, \tilde{q}_2+\varepsilon_n, q_3, \dots, q_N), \quad \textrm{and } \lim_{n\rightarrow \infty} \varepsilon_n = 0. \nonumber
\end{equation}
As
\begin{eqnarray}
\frac{g(x,y_n)-g(x,y)}{y_n-y} & = & \frac{g(H_f(P), H_f(Q_{\varepsilon_n}))-g(H_f(P), H_f(Q))}{H_f(Q_{\varepsilon_n})-H_f(Q)} \nonumber \\
 & = & \frac{H_f(P\otimes Q_{\varepsilon_n})-H_f(P\otimes Q)}{H_f(Q_{\varepsilon_n})-H_f(Q)} \nonumber
\end{eqnarray}
and as
\begin{eqnarray}
&&\lim_{n\rightarrow \infty}\frac{H_f(P\otimes Q_{\varepsilon_n})-H_f(P\otimes Q)}{H_f(Q_{\varepsilon_n})-H_f(Q)} \nonumber \\
&& = \lim_{n\rightarrow \infty} \frac{\sum_{i=1}^N \left[ f(p_i\tilde{q}_1) - f(p_i(\tilde{q}_1-\varepsilon_n)) + f(p_i\tilde{q}_2)-f(p_i(\tilde{q}_2+\varepsilon_n))\right] }{f(\tilde{q}_1)-f(\tilde{q}_1-\varepsilon_n) + f(\tilde{q}_2) - f(\tilde{q}_2+\varepsilon_n)} \nonumber \\
&& = \lim_{n\rightarrow \infty} \sum_{i=1}^N p_i \left[ \frac{f(p_i\tilde{q}_1-\varepsilon_n p_i)-f(p_i\tilde{q}_1)}{-\varepsilon_n p_i} - \frac{f(p_i\tilde{q}_2+\varepsilon_n p_i)-f(p_i\tilde{q}_2)}{\varepsilon_n p_i}\right] \times \nonumber \\
&& \times \left( \frac{f(\tilde{q}_1-\varepsilon_n)-f(\tilde{q}_1)}{-\varepsilon_n} - \frac{f(\tilde{q}_2+\varepsilon_n)-f(\tilde{q}_2)}{\varepsilon_n}\right)^{-1} \nonumber \\
&& = (f'(\tilde{q}_1)-f'(\tilde{q}_2))^{-1}\sum_{i=1}^N p_i(f'(p_i\tilde{q}_1)-f'(p_i\tilde{q}_2)), \nonumber
\end{eqnarray}
we find that $\frac{\partial}{\partial y}g(x,y)$ exists and that
\begin{equation}
\frac{\partial}{\partial y}g(x,y) = (f'(\tilde{q}_1)-f'(\tilde{q}_2))^{-1}\sum_{i=1}^N p_i(f'(p_i\tilde{q}_1)-f'(p_i\tilde{q}_2)).
\end{equation}
But since $\tilde{q}_1$ and $\tilde{q}_2$ are the same for any choice of $y$, it is independent of this choice - i.e.
\begin{equation}
\frac{\partial}{\partial y}g(x,y) = B(x), \quad \textrm{for all } x,y\in ]a,b[.
\end{equation}
Completely analogously it is shown that the partial derivative with respect to $x$ exists and that
\begin{equation}
\frac{\partial}{\partial x}g(x,y) = A(y), \quad \textrm{for all } x,y\in ]a,b[.
\end{equation}
Since $a$ and $b$ were arbitrary, it follows that
\begin{equation}
g(x,y) = C + \alpha x + \beta y + \gamma xy, \quad x,y\in ]0, -f'(0)[,
\end{equation}
and the theorem follows by the continuity of $H_f$ on finite distributions - in particular we have for $P=(p,1-p)$ and $Q=(q,1-q)$, $0<p,q<1$,
\begin{eqnarray}
H_f(P) & = & \lim_{q\rightarrow 1}H_f(P\otimes Q) \nonumber \\
 & = & \lim_{q\rightarrow 1}\left[ C+ \alpha H_f(P) +\beta H_f(Q) + \gamma H_f(P)H_f(Q)\right] \nonumber \\
 & = & C + \alpha H_f(P), \nonumber
\end{eqnarray}
so $C=0$ and (after analogous argument for $\beta$) $\alpha = \beta = 1$, which shows the validity of the theorem for all $x,y\in [0,-f'(0)[$.
\end{proof}

Comparing theorem \ref{thm:composability} with the corollary of theorem \ref{thm:converse} we see that any $f$-entropy which is composable is indeed a Tsallis entropy.

\section{Discussion}
The condition of composability is important in thermodynamical applications of entropy, and we have shown here that any $f$-entropy which is composable must be a Tsallis entropy. Thus, for example, we see that attempts of 2-parameter generalizations of Tsallis entropy as in \cite{tsallis07, topsoe04} as well as many suggestions of alternate entropy functions such as the well known Kaniadikis\footnote{more generally, any linear kombination of Tsallis entropies that does not give a new Tsallis entropy.} entropy will not be able to satisfy the condition of composability. 

More generally, entropy is often defined as a function $H(P)=\varphi\left(-\sum_i f(p_i)\right)$, i.e. as a function of $f$-entropy, where $\varphi$ is positive and strictly increasing. In particular, $\varphi$ is bijective, and our theorem shows that if $H$ is composable, then it is in fact a function of some Tsallis entropy. A well known example is the Rényi entropy defined for $\alpha > 0$ by
\begin{equation}
H_{\alpha}(P) = \frac{1}{1-\alpha}\log \left( \sum_i p_i^{\alpha} \right). \nonumber
\end{equation}
It is an interesting question whether there exists one choice of transformation $\varphi$ which in some sense is better than any other choice. If so, one would expect that either Tsallis or Rényi entropy pops up. For this question, F. Topsøe suggested that his notion of game theoretical equilibrium might be relevant (see \cite{topsoe04} for details). In \cite{abe05} it is shown that the Rényi entropy lacks certain physically important properties that the Tsallis entropy posses.

\section*{Acknowledgments}
I am indebted to Flemming Topsøe for good advice, insightful discussions concerning both the present work and more general game and information theory, and not least for support. Also, Robert Niven for some helpful advice.

\end{document}